\documentclass[12pt,thmsa]{article}

\usepackage{cite}

\begin{document}

\title{Accurate calculation of resonances in multiple--well oscillators}
\author{Francisco M. Fern\'andez, \\
INIFTA (Conicet, UNLP), Divisi\'on Qu\'imica Te\'orica, \\
Diag 113 S/N, Sucursal 4, Casilla de Correo 16, \\
1900 La Plata, Argentina, \\
E--mail: fernande@quimica.unlp.edu.ar}
\maketitle

\begin{abstract}
Quantum--mechanical multiple--well oscillators exhibit curious complex
eigenvalues that resemble resonances in models with continuum spectra. We
discuss a method for the accurate calculation of their real and imaginary
parts.
\end{abstract}

\section{Introduction \label{sec:intro}}

Some time ago, Benassi et al \cite{BGG79} discussed the occurrence of
complex eigenvalues, or ''resonances '', in some quantum--mechanical
multiple--well oscillators, and calculated them for a particular example.
Recently Killingbeck \cite{K07} showed that the Hill--series method yields
quite accurate results for both the real and imaginary parts of those
eigenvalues if one introduces a complex parameter in the exponential factor
of the expansion. In principle, one has to tune up this parameter in order
to obtain an acceptable rate of convergence. Such ''complexification'' of
the well--known Hill--series method had been tried successfully before in
perturbation and matrix approaches \cite{KGJ04,KGJ05,KGJ06}.
Complexification is a term coined to indicate the use of, for example, a
complex frequency in the treatment of a perturbed harmonic oscillator or a
complex atomic number in the case of a perturbed Coulomb problem \cite
{K07,KGJ04,KGJ05,KGJ06}.

The Riccati--Pad\'{e} method (RPM) is known to be suitable for the accurate
calculation of bound states and resonances of simple quantum--mechanical
models \cite{FMT89a,FMT89b,F92,FG93,F95,F95c,F96,F96b,F97}. However, it has
only been applied to the most commonplace resonances in the continuum
spectrum\cite{F95,F95c,F96,F96b,F97}. The purpose of this paper is to
investigate if the RPM is also a reasonable alternative to the calculation
of the unusual kind of resonances considered by Benassi et al\cite{BGG79}
and Killinbeck\cite{K07}.

In Section~\ref{sec:RPM} we outline the RPM and in Section~\ref{sec:Results}
we apply it to the three--well oscillator treated explicitly by Benassi et al%
\cite{BGG79} and Killingbeck\cite{K07}.

\section{The Riccati--Pad\'{e} method (RPM) \label{sec:RPM}}

In order to make this paper reasonably self--contained, in this section we
outline the RPM in a quite general way. Suppose that a solution to the
eigenvalue equation
\begin{equation}
\Psi ^{\prime \prime }(x)+\left[ E-V(x)\right] \Psi (x)=0
\label{eq:Schrodinger}
\end{equation}
can be expanded in the form
\begin{equation}
\Psi (x)=x^{\alpha }\sum_{j=0}^{\infty }c_{j}x^{\beta j},\;\alpha ,\beta >0.
\label{eq:Psi-series}
\end{equation}
The power--series expansion for the regularized logarithmic derivative
\begin{equation}
f(x)=\frac{\alpha }{x}-\frac{\Psi ^{\prime }(x)}{\Psi (x)}=x^{\beta
-1}\sum_{j=0}^{\infty }f_{j}x^{\beta j}  \label{eq:f-series}
\end{equation}
converges in a neighbourhood of $x=0$ and the coefficients $f_{j}$ depend on
the eigenvalue $E$. The function $f(x)$ is a solution to the Riccati
equation
\begin{equation}
f^{\prime }(x)-f(x)^{2}+\frac{2\alpha }{x}f(x)+V(x)-E-\frac{\alpha (\alpha
-1)}{x^{2}}=0.  \label{eq:Riccati}
\end{equation}

Equations (\ref{eq:Schrodinger})--(\ref{eq:Riccati}) apply to both
one--dimensional ($-\infty <x<\infty $) and central--field ($0\leq x<\infty $%
) models. If $V(x)$ is a parity--invariant one--dimensional potential, then $%
\alpha =0$ for even states, $\alpha =1$ for odd ones, and $\beta =2$ for
both cases. If $\lim_{x\rightarrow 0^{+}}x^{2}V(x)=$ $V_{-2}>0$, then $%
\alpha (\alpha -1)=V_{-2}$  removes the singularity at origin in the case of
a central--field model.

The RPM consists of rewriting the partial sums of the power series (\ref
{eq:f-series}) as  Pad\'{e} approximants $x^{\beta -1}[N+d/N](z)$, $%
z=x^{\beta }$, in such a way that
\begin{equation}
\lbrack N+d/N](z)=\frac{\sum_{j=0}^{N+d}a_{j}z^{j}}{\sum_{j=0}^{N}b_{j}z^{j}}%
=\sum_{j=0}^{2N+d+1}f_{j}z^{j}+\emph{O}(z^{2N+d+2}).  \label{eq:Pade}
\end{equation}
In order to satisfy this condition the Hankel determinant $H_{D}^{d}$, with
matrix elements $f_{i+j+d+1}$, $i,j=0,1,\ldots ,N$, vanishes, where $%
D=N+1=2,3,\ldots $ is the determinant dimension, and $d=0,1,\ldots $ is a
displacement\cite{FMT89a,FMT89b,F92,FG93,F95,F95c,F96,F96b,F97}. The main
assumption of the RPM is that there is a sequence of roots $E^{[D,d]}$ of
the Hankel determinants $H_{D}^{d}$ that converges towards a given
eigenvalue of the Schr\"{o}dinger equation (\ref{eq:Schrodinger}) as $D$
increases\cite{FMT89a,FMT89b,F92,FG93,F95,F95c,F96,F96b,F97}.\ For brevity
we call it a Hankel sequence.

Notice that one obtains the coefficients $f_{j}$ from the expansion of the
Schr\"{o}dinger equation (\ref{eq:Schrodinger}) or the Riccati equation (\ref
{eq:Riccati}) quite easily, and that unlike the Hill--series method \cite
{K07} the RPM does not require an adjustable complex parameter. Besides, it
is not necessary to take into account the boundary conditions explicitly in
order to apply the RPM, and, for that reason, the method provides both bound
states and resonances simultaneously\cite
{FMT89a,FMT89b,F92,FG93,F95,F95c,F96,F96b,F97}.

\section{Results and discussion \label{sec:Results}}

In what follows we apply the RPM to calculate the curious complex eigenvalue
of the triple--well oscillator
\begin{equation}
V(x)=x^{2}-2g^{2}x^{4}+g^{4}x^{6}  \label{eq:V(x)}
\end{equation}
reported by Benassi et al\cite{BGG79} and Killingbeck\cite{K07}. In this
case $\beta =2$ and we choose $\alpha =0$ for even states as discussed above.

Table~\ref{tab:convergence} shows a Hankel sequence $E^{[D,0]}$ that
converges towards the lowest complex eigenvalue when $g=0.14$. We have kept
twenty digits in all entries in order to show how they become stable as $D$
increases. Notice the remarkable rate of convergence of the Hankel sequence
for both the real and imaginary parts of the eigenvalue.

Table~\ref{tab:E(g)} shows the same complex eigenvalue for a range of $g$%
--values somewhat wider than the ones chosen by Benassi et al\cite{BGG79}
and Killingbeck \cite{K07}. We have truncated the results, obtained from
Hankel determinants with $D\leq 15$ and $d=0$, to the apparently last stable
digit. The first digits of our results agree with those given by Benassi et
al\cite{BGG79} and Killingbeck\cite{K07}. We notice that ${\rm Im}
E(g^{2})g^{2}\exp (1/(2g^{2}))$ does not seem to approach a constant for
those values of $g$. It may be that ${\rm Im}E(g^{2})$ attains the WKB
asymptotics\cite{BGG79} at smaller values of $g$.

It is interesting to compare the strange resonance of the potential (\ref
{eq:V(x)}) with the more commonplace one of the potential
\begin{equation}
V_{2}(x)=x^{2}-2g^{2}x^{4}  \label{eq:V2(x)}
\end{equation}
that was treated earlier by means of the RPM\cite{F95}. Table \ref{tab:E(g)2}
shows the lowest resonance for this model for the same values of $g$
considered before. We appreciate that the imaginary part of this resonance
is considerably greater than the previous one and that it seems to approach
the WKB asymptotics ${\rm Im}E^{WKB}(g^{2})=[4/(2\pi g^{2})]\exp
(-1/[3g^{2}])$ somewhat faster.

The results of this paper clearly show that the RPM is suitable for the
calculation of both real and complex eigenvalues of simple Hamiltonian
operators, even in the case of quite small imaginary parts. We believe that
this approach is a most useful tool in the numerical investigation of a wide
variety of eigenvalue problems. Its main advantages are: great rate of
convergence and simple straightforward application that does not require
adjustable parameters or explicit consideration of boundary conditions.

\begin{table}[]
\caption{Convergence of a Hankel sequence $E^{[D,0]}$ towards the lowest
complex eigenvalue of the oscillator (\ref{eq:V(x)}) with $g=0.14$.}
\label{tab:convergence}
\begin{center}
\begin{tabular}{rll}
\hline
\multicolumn{1}{c}{$D$} & \multicolumn{1}{c}{${\rm Re} E$} &
\multicolumn{1}{c}{${\rm Im} E$} \\ \hline
2 & $0.96913474062929793208$ & $0$ \\
3 & $0.96912933030952144688$ & $0$ \\
4 & $0.96912932029284635448$ & $0$ \\
5 & $0.96912932006642961226 $ & $3.6781221743857153252\ 10^{-10}$ \\
6 & $0.96912932002647227146 $ & $3.3990326234127550889\ 10^{-10}$ \\
7 & $0.96912932002710973379 $ & $3.3801038698293392418\ 10^{-10}$ \\
8 & $0.96912932002717289039 $ & $3.3798079586780234680\ 10^{-10}$ \\
9 & $0.96912932002717518442 $ & $3.3798093143407212241\ 10^{-10}$ \\
10 & $0.96912932002717525409 $ & $3.3798095397280767486\ 10^{-10}$ \\
11 & $0.96912932002717525622 $ & $3.3798095479442123313\ 10^{-10}$ \\
12 & $0.96912932002717525629 $ & $3.3798095481219295624\ 10^{-10}$ \\
13 & $0.96912932002717525629 $ & $3.3798095481219029216\ 10^{-10}$ \\
14 & $0.96912932002717525629 $ & $3.3798095481216587093\ 10^{-10}$ \\
15 & $0.96912932002717525629 $ & $3.3798095481216435223\ 10^{-10}$%
\end{tabular}
\par
\end{center}
\end{table}

\begin{table}[]
\caption{Complex eigenvalue of the oscillator (\ref{eq:V(x)}) for several
values of $g$.}
\label{tab:E(g)}
\begin{center}
\begin{tabular}{rllc}
\hline
\multicolumn{1}{c}{$g$} & \multicolumn{1}{c}{${\rm Re} E(g^2)$} &
\multicolumn{1}{c}{${\rm Im} E(g^2)$} & \multicolumn{1}{c}{${\rm Im}
E(g^2)\ g^2 \exp(1/(2g^2))$} \\ \hline
0.08 & $0.99025645954150600314 $ & $1.16994\ 10^{-32}$ & 0.6362094894 \\
0.09 & $0.98761765110834730415 $ & $1.28623698 \ 10^{-25}$ & 0.6700502315 \\
0.10 & $0.98464158830285882643 $ & $1.3513930260 \ 10^{-20} $ & 0.7006574893
\\
0.12 & $0.97763491479323529157 $ & $4.3530125379031 \ 10^{-14} $ &
0.7530467190 \\
0.14 & $0.96912932002717525629 $ & $3.37980954812164 \ 10^{-10} $ &
0.7944913345 \\
0.16 & $0.95896997046169207832 $ & $1.0619001732959989 \ 10^{-7} $ &
0.8253492417 \\
0.18 & $0.94691604067745932355 $ & $5.18077667159013113 \ 10^{-6} $ &
0.8453084682 \\
0.20 & $0.93255571582477452180 $ & $7.94775543996767651 \ 10^{-5} $ &
0.8530716514 \\
0.22 & $0.91525354748034208273 $ & $5.70253065914296141 \ 10^{-4} $ &
0.8461088416 \\
0.24 & $0.89442055320991452496 $ & $2.424632840047890532 \ 10^{-3} $ &
0.8222158493 \\
0.26 & $0.87011531157430539225 $ & $7.104058338260953225 \ 10^{-3} $ &
0.7828715436 \\
0.28 & $0.84333442392342060412 $ & $1.5915859465250206010 \ 10^{-2} $ &
0.7343132667 \\
0.30 & $0.81560795814733914293 $ & $2.9400216892153485663 \ 10^{-2} $ &
0.6844475376
\end{tabular}
\par
\end{center}
\end{table}

\begin{table}[]
\caption{Lowest resonance of the oscillator (\ref{eq:V2(x)}) for several
values of $g$.}
\label{tab:E(g)2}
\begin{center}
\begin{tabular}{rlll}
\hline
\multicolumn{1}{c}{$g$} & \multicolumn{1}{c}{${\rm Re} E(g^2)$} &
\multicolumn{1}{c}{${\rm Im} E(g^2)$} & \multicolumn{1}{c}{${\rm Im}
E(g^2)\ g \exp(1/(3g^2))$} \\ \hline
0.08 & $0.99017315154568105030 $ & $4.66667951 \ 10^{-22}$ &
1.554541174 \\
0.09 & $0.98748105548308533216 $ & $2.3014736620 \ 10^{-17}$ &
1.543296673 \\
0.10 & $0.98442766976525540084 $ & $5.1093948883947 \ 10^{-14}$ &
1.530566484 \\
0.12 & $0.97716020191841551216 $ & $1.1063680213861671 \ 10^{-9}$ &
1.500354438 \\
0.14 & $0.96816424784205963513 $ & $4.297124100601175228 \ 10^{-7}$ &
1.463074727 \\
0.16 & $0.95708500653988706061 $ & $1.9606870293524100682 \ 10^{-5} $ &
1.417112487 \\
0.18 & $0.94328218799381038166 $ & $2.5699864836055797687 \ 10^{-4}$ &
1.35910675 \\
0.20 & $0.92594246107314318252 $ & $1.5440221243204925966 \ 10^{-3}$ &
1.284707315 \\
0.22 & $0.90482508551985951067 $ & $5.5395017058573660278 \ 10^{-3}$ &
1.193719284 \\
0.24 & $0.88093011197386366807 $ & $1.3978475279423154843 \ 10^{-2}$ &
1.093828654 \\
0.26 & $0.85613353763295142744 $ & $2.767004146177769213 \ 10^{-2}$ &
0.9964939951 \\
0.28 & $0.83225989985769363726 $ & $4.6300611971065823176 \ 10^{-2}$ &
0.9104055713 \\
0.30 & $0.81052712217939364397 $ & $6.8908503646837670242 \ 10^{-2}$ &
0.839251556
\end{tabular}
\par
\end{center}
\end{table}

\end{document}